\documentclass[aps,prl,amssymb,showpacs,twocolumn]{revtex4}
\usepackage{graphicx}

\newcommand{\Case}[2]{{\textstyle \frac{#1}{#2}}}
\newcommand{\lP}{\ell_{\mathrm P}}

\begin{document}

\preprint{IMSc/2005/4/12}

\title{Pre-classical solutions of the vacuum Bianchi I loop quantum cosmology}

\author{Ghanashyam Date}
\email{shyam@imsc.res.in}
\affiliation{The Institute of Mathematical Sciences,
CIT Campus, Chennai-600 113, INDIA}


\begin{abstract}

Loop quantization of diagonalized Bianchi class A models, leads to a
partial difference equation as the Hamiltonian constraint at the quantum
level.  A criterion for testing a viable semiclassical limit has been
formulated in terms of existence of the so-called pre-classical
solutions. We demonstrate the existence of pre-classical solutions of
the quantum equation for the vacuum Bianchi I model.  All these
solutions avoid the classical singularity at vanishing volume.

\end{abstract}

\maketitle

Loop Quantum Gravity (LQG) is the leading candidate for a manifestly
background independent approach to constructing a quantum theory of
gravity \cite{LQGRev}. This approach is particularly well suited in the
context where the Einstein theory indicates occurrence of singularities
entailing highly dynamical geometries with extreme curvatures.  The
methods employed in this approach can be adapted and tested in the
simpler context of cosmological models. Quantizing the cosmological
models along the lines of LQG has lead to the development of Loop
Quantum Cosmology (LQC) \cite{LQCRev}. 

One of the crucial simplification available in LQC is the existence of
the triad representation and knowledge of complete spectrum of the
volume operator so crucial for quantization of the Hamiltonian
constraint. The fact that the holonomies of the connection are well
defined operators but not the connection it self, is directly
responsible for the two main features of LQC: (a) the quantum
Hamiltonian constraint leads to a {\em difference equation} \cite{LQCIV}
and (b) inverses of triad components have {\em bounded} spectra
\cite{InvScale,ICGCInv}. Both these features lead to the absence of
`singularities' in the quantum theory \cite{Sing,Bohr,HomCosmo,Spin}. 

While loop quantization of cosmological models is well specified, one
also needs to check if the quantum dynamics admits solutions
(semiclassical states) which can approximate the classical description.
A natural way to recover classical behavior would be in terms of the
expectation values of suitable observables in the semiclassical states.
LQC (and LQG) being constrained systems, makes such a recovery of
classical behavior, more complicated. The solutions of the Hamiltonian
constraint, the only relevant one for LQC, are typically distributional
and one needs to equip the space of solutions with a new (physical)
inner product. One also needs to identify suitable (Dirac) observables.
Addressing these aspects is at a preliminary stage
\cite{Golam,NPV,Dittrich,ABC}.

Current understanding of the semiclassical limit of LQC is centered
around the notion of {\em pre-classicality}.  The articulation of this
notion has undergone a few changes and it is useful to note them.
Originally proposed in \cite{DynIn}, pre-classicality was thought in
terms of obtaining the continuum Wheeler-De Witt equation from the
fundamental difference equation by a limit in which the Barbero-Immirzi
parameter is taken to zero. Subsequently, the idea of {\em pre-classical
limit} was replaced by the idea of a {\em pre-classical approximation}
\cite{FundamentalDiffEqn}. To obtain (modified) Einstein dynamics from
the Hamilton-Jacobi equation provided by a WKB approximation
\cite{SemiClass,EffHamiltonian,DiscretenessCorrections}, it was shown to
be sufficient to have {\em approximate} pre-classical solution(s) with a
finite domain of validity. The property of {\em local stability} is
needed in the construction of such solutions \cite{FundamentalDiffEqn}
and is satisfied by difference equations of LQC \cite{Spin}.  Recently,
the methods of analyzing the asymptotic behavior of exact solutions of
the difference equations have been developed
\cite{GenFunction,AsymPropIso} which provide the sharpest yet
formulation of pre-classicality of a solution \cite{GenFunction}.
Briefly, the criterion is that asymptotically, small scale {\em
oscillations} in the solution be suppressed.  

Pre-classical solutions so identified are known to exist for isotropic
models \cite{Sing,DynIn,AsymPropIso} and some of the LRS models
\cite{GenFunction}. However, for the anisotropic, vacuum Bianchi I
model, pre-classical solutions were shown not to exist
\cite{CartinKhanna}. While this result is true for solutions that {\em
pass through vanishing volume}, there are more possibilities which
permit existence of pre-classical solutions.  Since the works in
\cite{HomCosmo,Spin}, the so called Bohr quantization has been developed
\cite{Bohr} which is crucial for the existence of pre-classical
solutions and we incorporate it in the brief summary of the quantum
theory given below.

The kinematical Hilbert space is spanned by orthonormalized vectors
labeled as $|\mu_1, \mu_2, \mu_3 \rangle, ~\mu_I \in \mathbb{R}$. These
are properly normalized eigenvectors of the triad operators $p^I$ with
eigenvalues $\Case{1}{2} \gamma \lP^2 \mu_I$, where $\gamma$ is the
Barbero-Immirzi parameter and $\lP^2 := 8 \pi G \hbar := \kappa \hbar$.
The volume operator is also diagonal in these labels with eigenvalues
$V(\vec{\mu})$ given by $(\Case{1}{2}\gamma\lP^2)^{3/2} \sqrt{|\mu_1
\mu_2 \mu_3|}$.  Here we have used the vector notation to denote the
triple $(\mu_1, \mu_2, \mu_3)$.  Imposing the Hamiltonian constraint
operator on general vectors of the form $|s\rangle = \sum_{\vec{\mu}}
s(\vec{\mu}) |\vec{\mu}\rangle$ leads to the fundamental difference
equation for the wave function $s(\vec{\mu})$. Here the sum is over
countable subsets of $\mathbb{R}^3$. The wave function $s(\vec{\mu})$
have to be invariant under simultaneous reversal of signs of a pair of
$\mu_I$'s and corresponds to the residual freedom of reversing the sign
of any two of the triad components \cite{HomCosmo}. 

In the present context of vacuum Bianchi I model, the Hamiltonian
constraint leads to the equation \cite{Spin},
\begin{equation}\label{DiffEqnOne} 
\sum_{\vec{\epsilon}_{12}} A_{12}(\vec{\mu}; \vec{\epsilon}_{12})
s(\vec{\mu}; \vec{\epsilon}_{12})  + \text{cyclic} = 0\ , \text{where}
\end{equation}
$\vec{\epsilon}_{12} = (\epsilon_1, \epsilon_2, \epsilon'_1,
\epsilon'_2)$ with each of the $\epsilon_*$ taking values $\pm1$;
$s(\vec{\mu}; \vec{\epsilon}_{12}) = s(\mu_1 - \mu_0 \epsilon_1 - \mu_0
\epsilon'_1, \mu_2 - \mu_0 \epsilon_2 - \mu_0 \epsilon'_2, \mu_3)$;
$\mu_0$ is an order 1 parameter and,
\begin{eqnarray} \label{Defns}
A_{12}(\vec{\mu}; \vec{\epsilon}_{12}) & = &
V(\vec{\mu}; \vec{\epsilon}_{12}) d(\mu_3)( \epsilon_1 \epsilon_2 +
\epsilon'_1
\epsilon'_2) \\
d(\mu) & :=  & \left\{\begin{array}{cl} 
|1 + \mu_0\mu^{-1}|^{\frac{1}{2}} - |1 -
\mu_0\mu^{-1}|^{\frac{1}{2}}  &  
\mu \ne 0 \\
0 & \mu = 0 
\end{array} \right. \nonumber
\end{eqnarray}

Note that while the equation (\ref{DiffEqnOne}) is defined for all
$\mu_I \in \mathbb{R}$, it is actually a {\em difference} equation since
only the coefficients $s(\vec{\mu})$, differing in steps of $\pm 2
\mu_0$ are constrained by the equation. One can make this explicit by
setting $\mu_I := 2 \mu_0 \nu_I + 2\mu_0 N_I, N_I \in \mathbb{Z}$ and
$S_{\vec{N}}(\vec{\nu}) := s(\vec{\mu})$. Clearly, there are infinitely
many `sectors' labeled by $\vec{\nu}$ with $\nu_I \in [0, 1)$.  Only
those sectors for which {\em at least one} of the $\nu_I$ is zero, one
will encounter zero volume.

It is convenient to absorb the factors of volume eigenvalues into the
wave functions by defining $t(\vec{\mu}) := V(\vec{\mu}) s(\vec{\mu})$
so that the equation (\ref{DiffEqnOne}) becomes,
\begin{equation}\label{DiffEqnTwo}
\sum_{\vec{\epsilon}_{12}} d(\mu_3)( \epsilon_1 \epsilon_2 +
\epsilon'_1 \epsilon'_2) t(\vec{\mu};\vec{\epsilon}_{12}) +
\text{cyclic} = 0
\end{equation}
This preserves the gauge invariance condition on the $s(\vec{\mu})$.
Furthermore, due to the explicit volume eigenvalues, $t_{\mu_1, \mu_2,
\mu_3} = 0$ if any of the $\mu_I$'s equal zero.

Making a product ansatz for the wave function and introducing the
difference operator $\Delta$, 
\begin{eqnarray}\label{Ansatz}
t(\vec{\mu}) & := & z_1(\mu_1)\ z_2(\mu_2)\ z_3(\mu_3), \\
\Delta z_I(\mu_I) & := & \left\{z_I(\mu_I + 2
\mu_0) - z_I(\mu_I - 2 \mu_0)\right\},
\end{eqnarray}
the difference equation (\ref{DiffEqnTwo}) can be written as,
\begin{equation}\label{DiffEqnProduct}
d(\mu_3)\Delta z_1(\mu_1) \Delta z_2(\mu_2) z_3(\mu_3)
+ \text{cyclic} = 0 \ .
\end{equation}
The gauge invariance conditions then translate to $z_I(- \mu_I) = \eta
z_I(\mu_I), \forall I$, where $\eta = \pm 1$. The vanishing of
$t(\vec{\mu})$  when any of the $\mu_I = 0$, translates into the
condition: $z_I(0) = 0, \forall I$. 

An exact solution of the partial difference equation can be obtained by
setting,
\begin{equation}\label{zeq}
\Delta z_I(\mu_I) ~ = ~ \beta_I d(\mu_I) z_I(\mu_I)~,~~ \forall I~,~~
\forall \mu_I \in \mathbb{R}
\end{equation}

where $\beta_I$ are some constants which have to satisfy,
\begin{equation} \label{KasnerCond}
\beta_1 \beta_2 + \beta_2 \beta_3 + \beta_3 \beta_1 ~ = ~ 0 
~ = ~ (\sum_I \beta_I)^2 - \sum_I \beta_I^2 \ .
\end{equation}

Thus we obtain a class of exact solutions of the {\em partial}
difference equation (\ref{DiffEqnTwo}) from those of three, {\em
ordinary} difference equations (\ref{zeq}) with parameters $\beta_I$
satisfying (\ref{KasnerCond}).

The original partial difference eq. (\ref{DiffEqnOne}) is {\em linear}
and with {\em real} coefficients, $A_{IJ}$'s. So, {\em without loss of
generality}, we can assume the wave function, $s(\vec{\mu})$ to be real.
Since for the product ansatz, all $z_I$ are independent, these must be
real as well which requires that the $\beta_I$'s be real. 

If all $\beta_I$ are zero, then $z_I$'s are constants and so is
$t(\vec{\mu})$. These {\em cannot} satisfy the condition $z_I(0) = 0$
without making $t(\vec{\mu}) = 0$ identically. The non-trivial solutions
then cannot pass through zero volume and must belong to the sectors with
$\nu_I \ne 0$. This also shows that in the sectors with $\nu_I \ne 0,
\forall I$, there {\em is an exact pre-classical solution} to
(\ref{DiffEqnOne}) namely $t(2 \mu_0 \vec{\nu} + 2 \mu_0 \vec{N}) =
\text{constant} (\ne 0), \forall \vec{N} \in \mathbb{Z}^3$.  

If at least one $\beta_I \ne 0$, then we can always take out a common
factor from all $\beta_I$ and ensure $\sum_I \beta_I = 1$. Equivalently,
a common scaling of $\beta_I$ can be absorbed by a common inverse
scaling of the $d(\mu)$ functions, which amounts to a scaling of the
volume which cancels out in (\ref{DiffEqnOne}). The class of solutions
that is being constructed can thus be parameterized exactly in the same
manner as the classical Kasner solution. In particular either two of the
$\beta$'s are zero or exactly one is negative while other two are
positive.  From now on we will restrict to $0 < |\beta_I| < 1$.

To explore pre-classicality of the separable solution, let us focus on
(\ref{zeq}), suppressing the label $I$.  Introducing the notation: $\mu :=
2 \mu_0 \nu + 2 \mu_0 n, n \in \mathbb{Z}, \nu \in [0, 1)$ and $z(\mu)
:= Z^{(\nu)}_n$, (\ref{zeq}) can be written as:
\begin{eqnarray} \label{OdeZero}
Z^{\nu}_{n + 2} & = & Z^{\nu}_n  +  \beta d(\nu, n + 1) 
Z^{\nu}_{n + 1} ~~,~~\forall n \in \mathbb{Z} \\
d(\nu, n) & := & \left|1 + \frac{1}{2 (\nu + n)}\right|^{\frac{1}{2}}
- \left|1 - \frac{1}{2 (\nu + n)}\right|^{\frac{1}{2}} \nonumber 
\end{eqnarray}

We have infinitely many decoupled sectors, labeled by $\nu$ and for each
of these we have a {\em second order}, ordinary difference equation. Due
to linearity, only one condition is enough to determine a solution.
Only for the sector $\nu = 0$ ($\mu$ is integer multiple of $2 \mu_0$),
the condition $z(0) = 0$ is relevant and it fixes the solution
completely.  The gauge invariance condition translates into the
identification: 
\begin{equation}\label{GaugeInv}
Z^{1 - \nu}_{- (n + 1)} ~=~ \eta Z^{\nu}_n, ~~
\forall ~ n \in \mathbb{Z}, ~~ \eta = \pm 1
\end{equation}
which restricts the $Z$'s in the {\em same} sector, only for $\nu = 0,
\Case{1}{2}$.  For all other sectors the gauge invariance condition
relates {\em two different} sectors. Under the identification implied by
(\ref{GaugeInv}), the equation satisfied by $Z^{\nu}_n$ goes over to the
equation satisfied by $Z^{1 - \nu}_n$ automatically with the same value
of $\beta$.  Therefore the solutions in the sector $(1 - \nu)$ can all
be obtained from solutions in the sector $\nu$ via (\ref{GaugeInv}). For
$\nu = \Case{1}{2}$, the gauge invariance condition requires
$Z^{1/2}_{-1} = \eta Z^{1/2}_0$ which fixes the solution completely. For
the two sectors, $\nu = 0, \Case{1}{2}$, pre-classicality is not
optional -- the solution is either pre-classical or it is not. From now
on the superscript $\nu$ is suppressed. 

Consider (\ref{OdeZero}). Defining the ratios $u_{n + 1} := Z_{n +
1}/Z_n$, one can see that $[(u_{n + 2} - \beta d(\nu, n + 1)) u_{n + 1}
- 1]Z_n = 0$. There are several possibilities now ($n \ge 0$ for
definiteness). If $Z_n$ remains non-zero for all $n$, then $u_n \to \pm
1$ as $n \to \infty$. The $u_n \to -1$ is referred to as a sequence with
sign oscillations. Since $u_{n + 2} - \beta d(\nu, n + 1)$ and $u_{n +
1}$, must have the same sign, it is clear that if $u_1 > 0$ and $\beta >
0$, then $u_n > 0$ for all $n$.  For $\beta < 0$ however, $u_n$ may
become negative for some $n_0$ and then stay negative subsequently.
Whether this could happen depends on the value of $u_1$. Thus it is
conceivable that for some positive values of $u_1$, one could have a
sequence without sign oscillations.  If $Z_n$ converges to zero, then
$|u_n|$ converges to a value $\le 1$, once again allowing for $u_n \to
-1$.  The sequence $u_n$ could also converge to $0$ in which case there
could be oscillations about $0$, but these are suppressed.  By
pre-classicality we mean either absence or suppression of sign
oscillations. 

To identify pre-classical sequences, we employ the generating function
technique \cite{GenFunction}. The function $d(\mu)$ being an algebraic
function poses some difficulties which can be handled in exactly the
same way as in \cite{GenFunction}.  Basically, one separates out the
large $n$ part of $d(\nu, n)$ and solves the equation perturbatively,
$Z_n = a_n + \sum_{k = 1}^{\infty} Z^k_n$. The leading order term, $a_n$
satisfies the equation with $d(\nu, n)$ replaced by $(2(n + \nu))^{-1}$.
Let us focus on $n \ge 0$ so that the absolute signs can be removed.
Setting $\beta = 4 \lambda$ the equation defining the $a_n$ sequence is,
\begin{equation} \label{OdeOne}
a_{n + 2} - \frac{2 \lambda}{n + 1 + \nu} a_{n + 1} - a_n = 0 ~~,~~n
\ge 0 \ .
\end{equation}

To account for non-integral $\nu$, we define a generating function $F(x)
:= \sum_{n = 0}^{\infty} a_n x^{n + \nu}$ and the function $G(x) :=
x^{-1}(F(x) - a_0 x^{\nu})$, which satisfies a differential equation
equivalent to the difference equation (\ref{OdeOne}),
\begin{equation} \label{DifferentialEqn} 
\frac{d}{dx}( (1 - x^2)G ) - 2 \lambda G - 
a_0 (\nu + 1) x^{\nu} - a_1 \nu x^{\nu - 1}  = 0
\end{equation}

For non-zero $\nu$, the last term is singular at $x = 0$ implying that
$G(x)$ will not be analytic at $x = 0$.  However, the singular term is
integrable such that $G(x)$ is continuous at $x = 0$.  The series
representation requires it to vanish at $x = 0$.

The equation (\ref{DifferentialEqn}) can be easily integrated to give,
\begin{eqnarray} \label{GenSoln}
G(x) & = & (1 + x)^{\lambda - 1} (1 - x)^{-\lambda - 1}
\left[c_0 + \right. \nonumber \\
& & \left. \int^x \left(\frac{1 - t}{1 + t}\right)^{\lambda}
\left\{a_0(1 + \nu)t^{\nu} + a_1 \nu t^{\nu -1} \right\} \right]
\end{eqnarray}

Notice that given any $a_0, a_1$, the solution to (\ref{OdeOne}) is
completely determined and so should be $G(x)$. The above solution for
$G(x)$ contains an {\em indefinite} integral and an arbitrary constant
of integration, $c_0$. We must choose $c_0$ and convert the indefinite
integral to a definite one such that $G(x)$ corresponds to the sequence
specified by the given $a_0, a_1$. The only value of $G(x)$ we know
without having to know the full sequence is $G(0) = 0$. Furthermore both
the integrands in the integrals in (\ref{GenSoln}) are integrable at $x
= 0$. Thus it is possible to impose $G(0) = 0$ which then requires $c_0
= 0$. The generating function is then obtained as,
\begin{eqnarray}
(1 - x)G(x) & = & (1 + x)^{\lambda - 1} (1 - x)^{-\lambda} \times
\label{Soln}\\
& & \left[ a_0 I(\nu, \lambda, x) + a_1 I(\nu - 1, \lambda, x) \right] ,
\nonumber  \\
I(\nu, \lambda, x) & := & (1 + \nu)\int^x_0 \left(\frac{1 - t}{1 +
t}\right)^{\lambda} t^{\nu} \label{IntglDefn}
\end{eqnarray}
One may already note that the integrals are finite at both $x = \pm 1$
but the pre-factor is not. Thus singularities of $(1 - x)G(x)$ are
controlled by the pre-factor which is independent of $\nu$. Apart from
the $\nu$ dependence of the integrals, one does not expect qualitative
behavior of $G(x)$ to be affected by $\nu$. Furthermore, since there
are two free parameters ($a_0, a_1$) which specify the sequence and only
one of these is relevant one due to the linearity, one can at the most
impose only {\em one} condition capturing pre-classicality, to get a
non-trivial solution. 

Now consider the behavior of $G(x)$ as $x \to -1$. For the range of $0
< |\lambda| < 1$, both the integrals exist (and are positive), but the
pre-factor diverges. A divergence in $G(x)$ at $x = -1$ implies
un-suppressed sign oscillations which are to be avoided for
pre-classical sequences \cite{GenFunction}. Clearly, to avoid this
singularity in $G(x)$, the integrals must add up to zero which
determines $a_1$ in terms of $a_0$. For $x = -1$, one has,
\begin{eqnarray}\label{HyperGeom}
I(\nu, \lambda, -1) & = & (-1)^{\nu} I(\nu, -\lambda, 1)\ , \\
I(\nu, -\lambda, 1) & = & (1 + \nu) B(1 + \nu, 1 - \lambda) \times \nonumber
\\
& & 
F(-\lambda, \nu + 1, -\lambda + \nu + 2; -1) \ ,
\end{eqnarray}
where, $B$ and $F$ are the Beta function and the hypergeometric
functions \cite{HandBook}. The condition of no singularity at $x = -1$
gives,
\begin{equation}\label{NoOscillations}
a_0 I(\nu, -\lambda, 1) ~ = ~ a_1 I(\nu - 1, -\lambda, 1)\ .
\end{equation}

This determines the sequence {\em uniquely} modulo a trivial, overall
scaling.  The sequence satisfying (\ref{NoOscillations}) has suppressed
sign oscillations. Its convergence properties are determined by the $x
\to 1$ behavior of $(1 - x)G(x)$. For $x = 1$ the integrals again exist
but now the pre-factor {\em diverges for $\lambda > 0$} and {\em
vanishes for $\lambda < 0$} and so does the sequence $\{a_n\}$.
However, $(1 - x)G(x)$ is integrable at $x = 1$ which implies that the
asymptotic behavior of $a_n$ is bounded by $n$. Making a power law
ansatz for asymptotic $a_n$, one can see from (\ref{OdeOne}) (and indeed
from (\ref{OdeZero}) as well) that $a_n \sim n^{\lambda}$.  Since
$\beta_I$ come with both signs, both behaviors must be admissible. For
$\lambda = 0$, one gets $a_1 = a_0$ and the sequence is the constant
sequence $a_n = a_0 ~ \forall n \ge 0$ which is obviously pre-classical.

As anticipated, these results are exactly analogous to those obtained in
\cite{GenFunction}. Indeed, for $\nu = 0$, equations (\ref{OdeOne},
\ref{DifferentialEqn}, \ref{GenSoln}) go over to the equations of
\cite{GenFunction}. Now the boundary condition is $G(0) = a_1$ which
gives $c_0 = a_1$ and definite integral has the lower limit as $0$.
Demanding non-singularity of $G(-1)$, determines $a_1$ exactly as in
\cite{GenFunction}. The behavior at $x = 1$ is similar to that for the
non-zero $\nu$ case. 

Thus, there certainly exist sectors such that for each choice of the
separation constant $\beta, \ 0 < |\beta| < 1$, one can select {\em a
unique} solution of (\ref{OdeOne}) which is pre-classical.  These
solutions of course have to be improved by computing the corrections
$Z^k_n$ \cite{GenFunction}. The asymptotic power law behavior of $a_n$
will continue to hold also for $Z_n$. 

All these statements hold for $n \ge 0$. Having determined $Z_0^{\nu},
Z_1^{\nu}$, by pre-classicality, $Z_{n < 0}$ can be determined by the
{\em exact} equation (\ref{OdeZero}). Whether $Z_n^{\nu}$ is
pre-classical also for $n < 0$, can be inferred by testing for
pre-classicality of $Z_n^{1 - \nu}$ for positive $n$, using
(\ref{GaugeInv}). 

As noted earlier, the sectors $\nu = 0, \Case{1}{2}$ already have a
unique solution due to the conditions $Z^0_0 = 0$ and $Z^{1/2}_{-1} =
\eta Z^{1/2}_0$ respectively. If these conditions are imposed on the
leading pre-classical sequence $\{a_n\}$, then clearly there are {\em
no} non-trivial solutions in the $\nu = 0$ sector. For $\nu = 1/2$
sector, numerically, the gauge invariance condition and the
pre-classicality condition seem to hold only for $\beta = 0$ with $\eta
= 1$. For other sectors, pre-classicality is the only condition imposed
and solutions can be constructed.

The full wave function is the product of the three sequences and apart
from an over all constant factor, is completely determined.  One can
build more general (and non-separable) solutions by taking complex
linear combinations with coefficients being functions of $\vec{\beta}$.
Clearly, a combination involving a diverging and a vanishing solution
will be a diverging one and still without sign oscillations. Since such
a solution involves a distribution of $\vec{\beta}$, these parameters
themselves would not be identified with the classical Kasner parameters,
$\alpha_I$ (say) satisfying $\sum_I \alpha_I = 1 = \sum_I \alpha_I^2$.
Rather, one would imagine constructing linear combinations which `peak'
in some suitable sense, around three triad values $p^I_0$ and a Kasner
parameter $\vec{\alpha}_0$. (Since the reduced phase space of the vacuum
Bianchi I model is $4 = 6 -2$ dimensional, one needs {\em four}
parameters to specify a classical solution and these could be
conveniently taken as {\em three} initial triad values and a Kasner
parameter value.) If such a construction can be carried out, then one
would be able to claim that the quantum theory has `sufficient number of
semiclassical solutions' as expected from the classical theory. 

We note that since solutions of the Hamiltonian constraint are
expected to be distributional in general, kinematical normalizability of
the pre-classical (or otherwise) solutions is not directly mandated.
The requirement of pre-classicality for both signs of $n$ is an
open issue. For an alternative treatment of separable solutions, see
\cite{CartinKhannaSeparable}.

In summary, we have shown that in every sector $\nu_I \neq 0,
\Case{1}{2}$, there exist a one parameter family of pre-classical
solutions. For $\beta_I = 0$, the solution is in fact {\em exact},
possibly corresponding to the Minkowski space-time. All these solutions
skip the vanishing volume eigenvalues. By contrast, in the $\nu_I = 0$
sector, there are no pre-classical solutions \cite{CartinKhanna}.  The
richness of the loop quantization, manifested by the infinitely many
sectors, is crucial for this result; an observation also made in
\cite{NPV} in the isotropic context.  We have heuristically indicated
how these families can be used to see if loop quantization does admit
`enough semiclassical states'.  The exact, non-singular solutions of the
effective dynamics of vacuum Bianchi I models given in
\cite{NonSingKasner} also exhibit a similar feature of avoiding
vanishing volume which motivated this work. 

\begin{acknowledgments}
I would like to thank Martin Bojowald for helpful remarks particularly
regarding vanishing and diverging sequences vis a vis pre-classicality.
Discussions with Golam Hossain are also acknowledged. 
\end{acknowledgments}

\end{document}